\documentclass[aps,prl,twocolumn,superscriptaddress]{revtex4-1}
\usepackage{graphicx}
\usepackage{makecell}
\usepackage{multirow}
\usepackage{color}
\usepackage[dvipsnames]{xcolor}
\usepackage[colorlinks=true,urlcolor=Blue,linkcolor=Blue,citecolor=Blue]{hyperref}
\usepackage[all]{hypcap}
\usepackage{amsmath}
\usepackage{threeparttable,booktabs}
\usepackage[normalem]{ulem}

\newcommand{\minew}[1]{{\color{red}{#1}}}
\newcommand{\yl}[1]{{\color{blue}{#1}}}

\begin{document}

\preprint{APS/123-QED}

\title{Exploring isospin-nonconserving effects in the upper $fp$ shell \\with new mass measurements}

\author{H.F. Li}
\affiliation{CAS Key Laboratory of High Precision Nuclear Spectroscopy, Institute of Modern Physics, Chinese Academy of Sciences, Lanzhou 730000, China}
\author{X. Xu}%
 \email{xuxing@impcas.ac.cn}
\affiliation{%
CAS Key Laboratory of High Precision Nuclear Spectroscopy, Institute of Modern Physics, Chinese Academy of Sciences, Lanzhou 730000, China}%
\author{Y. Sun}%
 \email{sunyang@sjtu.edu.cn}
\affiliation{%
School of Physics and Astronomy, Shanghai Jiao Tong University, Shanghai 200240, China}%
\author{K. Kaneko}%
\affiliation{%
Department of Physics, Kyushu Sangyo University, Fukuoka 813-8503, Japan}%
\author{X. Zhou}%
\affiliation{%
CAS Key Laboratory of High Precision Nuclear Spectroscopy, Institute of Modern Physics, Chinese Academy of Sciences, Lanzhou 730000, China}%
\author{M. Zhang}%
\affiliation{%
CAS Key Laboratory of High Precision Nuclear Spectroscopy, Institute of Modern Physics, Chinese Academy of Sciences, Lanzhou 730000, China}%
\author{ W.J. Huang}%
\affiliation{%
Advanced Energy Science and Technology Guangdong Laboratory, Huizhou 516007, China}
\affiliation{%
CAS Key Laboratory of High Precision Nuclear Spectroscopy, Institute of Modern Physics, Chinese Academy of Sciences, Lanzhou 730000, China}%
\author{X.H. Zhou}%
\affiliation{%
CAS Key Laboratory of High Precision Nuclear Spectroscopy, Institute of Modern Physics, Chinese Academy of Sciences, Lanzhou 730000, China}%
\author{Yu.A. Litvinov}%
\affiliation{%
GSI Helmholtzzentrum f\"ur Schwerionenforschung, Planckstraße 1, 64291 Darmstadt, Germany}%

\author{M. Wang}%
\affiliation{%
CAS Key Laboratory of High Precision Nuclear Spectroscopy, Institute of Modern Physics, Chinese Academy of Sciences, Lanzhou 730000, China}%
\author{Y.H. Zhang}%
\affiliation{%
CAS Key Laboratory of High Precision Nuclear Spectroscopy, Institute of Modern Physics, Chinese Academy of Sciences, Lanzhou 730000, China}%



\date{\today}

\begin{abstract}
Nuclear mass measurements have recently been extended conspicuously to  proton-rich region in the upper $fp$ shell. 
\yl{The} 
new data \yl{are utilized to} study  isospin symmetry breaking \yl{phenomena} using Coulomb displacement energy (CDE) and triplet displacement energy (TDE) as probes. 
The new mass data, either measured for the first time or with greatly improved accuracy, removed several previously found ``anomalies"  \yl{in} the systematical behavior in the $fp$ shell. Remarkably, more regular odd-even staggering patterns can be established in both CDE and TDE, calling for a uniform explanation in terms of isospin-nonconserving (INC) forces across the $sd$, $f_{7/2}$, and upper $fp$ shells. By extending the large-scale shell-model calculation [Phys. Rev. Lett. \textbf{110}, 172505 (2013)] to the upper $fp$-shell region, we found that, in order to describe the new data, the same INC force is required as  previously used for the $f_{7/2}$ shell. 
Especially, we propose the $T=1$ TDE for those triplet nuclei, that have $pp$, $nn$, and $pn$ pairs on top of a common even-even $N=Z$ core, to be a good indicator for the isotensor component of  isospin violating interactions, which is estimated \yl{here} to be 150 keV. 

\end{abstract}

\maketitle


The nearly-equal masses of proton and neutron as well as the nearly-identical properties of nucleon-nucleon (NN) interactions lead to the elegant concept of isospin symmetry \cite{heisenberg1989,wigner1937}. 
In nuclei, the Coulomb interaction and  charge-dependent nucleon-nucleon interaction break this symmetry, giving rise to observable signatures. Especially for nuclei near the $N = Z$ line, effects of isospin-symmetry breaking (ISB) appear both in binding energies and \yl{excitation} spectra  \cite{bentley2007,kaneko2012,Baczyk2018}. \yl{Thus,} investigation of \yl{very exotic} proton-rich nuclei  provides important testing ground for ISB, thereby enhancing our knowledge on nuclear forces and nuclear many-body effects far from stability.

\minew {There are various probes to address isospin-nonconserving (INC) forces in atomic nuclei\,\cite{ormand1989,zuker2002}.
Commonly used quantities  are the mirror energy difference (MED) and triplet energy difference (TED), defined as excitation energy  shift between mirror (analogue) states.
Extensive experimental studies have been performed to explore MED between mirror pairs in the $sd$, $fp$ and $fpg$ shells~\cite{Jenkins2005,Vedova2007,kaneko2013,Ruotsalainen2013,Debenham2016,YAJZEY2021,FERNANDEZ2021,mihai2022,LLEWELLYN2020,Uthayakumaar2022}.
Numerous TEDs for $A = 46$\cite{garrett2021}, $A = 54$\cite{gadea2006}, $A = 62$ \cite{mihai2022}, $A = 66$\cite{Ruotsalainen2013} and $A = 70$\cite{Debenham2016}, $A = 74$ \cite{henderson2014} triplets were  thoroughly reviewed.
Inspired by these spectroscopic data, ISB effects on MEDs and TEDs were investigated by the large-scale shell-model calculations \cite{kaneko2014,lenzi2018}.

Another category of suitable quantities are Coulomb displacement energy (CDE) and triplet displacement energy (TDE) determined from the masses of the ground states.}
CDE and TDE are respectively proportional to the linear ($\propto T_z$) and quadratic ($\propto T_z^2$) terms of  the Wigner's Isobaric Multiplet Mass Equation (IMME) \cite{wigner1958proceedings}, where $T_z = (N-Z)/2$ is the $z$ component of isospin $T$ for a nucleus with $Z$ protons and $N$ neutrons. These terms are essentially related to the isovector and isotensor components of the charge violating interactions, and thus can potentially yield separate information on the charge symmetry and charge independence of the attractive $NN$ interaction \cite{henly1969, NN_sca_length_all,MACHLEIDT20111}.

Detailed knowledge on CDE and TDE has further far-reaching implications in fundamental physics. 
For example, 
super-allowed $0^+ \rightarrow 0^+$ $\beta$ decays yield the most precise value for $V_{\rm ud}$, the up-down element of the Cabibbo-Kobayashi-Maskawa (CKM) matrix, which is the key ingredient to test the CKM-matrix unitarity, a fundamental requirement of the electroweak standard model. To extract $V_{\rm ud}$ from experimental $ft$ data, theoretical corrections must be applied to take account of unobserved ISB effects in the analog states of the parent and daughter nuclei \cite{ormand1995,towner2008}. For this purpose, experimental CDE and TDE \cite{britz1998} serve as a calibration for nuclear shell models utilized to  calculate such corrections \cite{towner2008,kaneko2017}. Reference \cite{kaneko2017} showed that anomalies observed in super-allowed $0^+ \rightarrow 0^+$ $\beta$ decay can only be explained by introducing INC forces into the model, which is achieved via fitting the experimental CDEs and TDEs. According to the 2020 compilation of super-allowed $0^+ \rightarrow 0^+$ $\beta$ decays \cite{hardy2020}, these transitions are now known in the decays of nuclei up to $^{74}$Rb in the upper $fp$ shell, for which the ISB is the discussion topic of the present work.

Another important \yl{implication} is related to the comprehension of the stellar nucleosynthesis processes\,\cite{mumpower2016, Yamaguchi-2021}.
Precise masses especially in the vicinity of the waiting-point nuclei \cite{schatz1998} are needed to locate the $rp$-process path 
and to infer properties of the underlying accreting neutron stars\,\cite{Meisel-2019}.
The CDE has been thought to be a reliable quantity for predictions of unknown masses on the proton-rich side of the $N=Z$ line \cite{Ormand1997}.  
Using CDE, one can not only derive  nuclear masses, but also determine positions of unknown resonances, provided that level schemes of neutron-rich mirror nuclei are experimentally known \cite{herndl1995,fisker2001}. Moreover, using TDE one can obtain precious information for $T_z = -1$ nuclei if experimental values exist for $T_z = 0$ and $T_z=+1$ nuclei belonging to the same $T = 1$ multiplets \cite{richter2011,brown2014}.

The CDE for mirror nuclei, defined as 
\begin{equation}\label{eq:cde}
	{\rm CDE}(A,T) = {\rm BE}(T,T_{z<}) - {\rm BE}(T,T_{z>}),
\end{equation}
is an established measure of charge-symmetry breaking ~\cite{bentley2007}, where BE$(T,T_{z<}$) and BE$(T,T_{z>}$) are binding energies of the mirror partners having the smaller proton number $(Z_<$) and larger proton number $(Z_>$), respectively. Under the assumption of spherical nuclei, the CDE exhibits a linear dependence on $Z/A^{1/3}$ regardless of the shell structure ~\cite{osti_5713315,ANTONY19971,miernik2013mass}.
This relationship could be taken as an initial estimate for CDE within an uncertainty of about 150~keV~\cite{sunplb2020}. Hartree–Fock (HF) calculations could reproduce the experimental CDEs within an uncertainty of about 100~keV.
Masses of proton-rich nuclei  \yl{obtained from such} calculated CDEs \yl{combined} with known masses of the neutron-rich mirror partners have been used for mapping the proton drip-line and in the $rp$-process calculations~\cite{brown2002}.
Influence of quadruple deformation on CDEs was also studied  ~\cite{antony1986coulomb,tu2014}. It is well known that due to the strong Coulomb repulsion in systems with all protons paired, CDEs exhibit odd-even staggerings with $Z$, which leads to larger CDEs for the mirror pairs having  odd-$Z_>$ and even-$Z_<$~\cite{Feenberg1946}. To emphasize the staggerings, $\Delta$CDE is defined between \yl{nuclei with mass numbers}  $A$ and $A + 2$,
\begin{equation}\label{eq:dcde}
	\Delta {\rm CDE}(A,T) = {\rm CDE}(A+2,T) - {\rm CDE}(A,T).
\end{equation}
The TDE for $T = 1$ triplets, a signature of charge-independence breaking, is defined as
\begin{equation}\label{eq:tde}
\begin{aligned}
	{\rm TDE}(A,T) &= {\rm BE}(T,T_{z<}) + {\rm BE}(T,T_{z>})\\
	&- 2{\rm BE}(T, T_z = 0).
\end{aligned}
\end{equation}
To magnify the trend with increasing $Z$, $\Delta$TDE for nuclei differing in mass number by four is defined as
\begin{equation}\label{eq:dtde}
	\Delta {\rm TDE}(A,T) = {\rm TDE}(A+4,T) - {\rm TDE}(A,T).
\end{equation}

In the 2013 work \cite{kaneko2013}, two of us (KK and YS) and collaborators carried out an investigation on the INC effects for masses $A= 42-95$ using CDEs and TDEs as probes. Large-scale shell-model
calculations were performed by employing two modern
effective interactions (GXPF1A and JUN45) for the corresponding mass regions with inclusion of the Coulomb plus INC nuclear interaction. The INC interaction was applied only in the $f_{7/2}$ shell to understand the observed \yl{staggering} in $\Delta$CDE~\cite{kaneko2013}. A clear conclusion was made that the INC forces are important for the $f_{7/2}$-shell nuclei.

However, no conclusion about INC forces could be drawn for the upper $fp$-shell and heavier nuclei~\cite{kaneko2013}. Even for the $f_{7/2}$ shell, some puzzles  remained. 
An apparent change in the staggering phase of the experimental $\Delta$CDE/$Z$ for $T = 1/2$ at $A = 69$ was left as an open question ~\cite{kaneko2013}. 
The \yl{calculations} could reproduce the smooth trend of the TDEs for $A = 4n + 2$ in lighter masses, where $n$ is an integer number, as well as a sudden decrease at $A = 58$ and the corresponding drop in $\Delta$TDE at $A = 54$. 
However, the TDE analysis for the sequence of $A = 4n + 2$ \yl{nuclei was} stopped at $A = 58$ because no experimental masses were available. Furthermore, the TDE of $A = 4n$ could not be determined due to the lack of experimental masses of $T_z = -1$ nuclei. 

Recently, plenty of experiments on mass \yl{measurements of} proton-rich nuclei have been performed. \yl{The masses were} either measured for the first time or with greatly improved accuracy. 
To mention a few,
masses of neutron-deficient gallium isotopes, $^{60-63}$Ga, were measured with the TITAN multi-reflection time-of-flight mass spectrometer~\cite{titan_mass}. At the experimental Cooler-Storage Ring (CSRe) in Lanzhou, masses of $T_z = -3/2$, $-1$, and $-1/2$ nuclei were measured~\cite{zhangmepja2023,wangmprl2023,zhouxnp2023} with high precision by an upgraded method of isochronous mass spectrometry (IMS). This novel technique,  named B$\rho$-defined IMS \cite{wangmprc2023}, greatly advanced mass measurements of neutron-deficient nuclei toward the heavier mass region, allowing for a considerable extension of our knowledge on  systematic behavior of CDE and TDE to the upper $fp$ shell. 

\begin{figure*}[htbp] \label{fig:fig1}
\includegraphics[width=1.7\columnwidth]{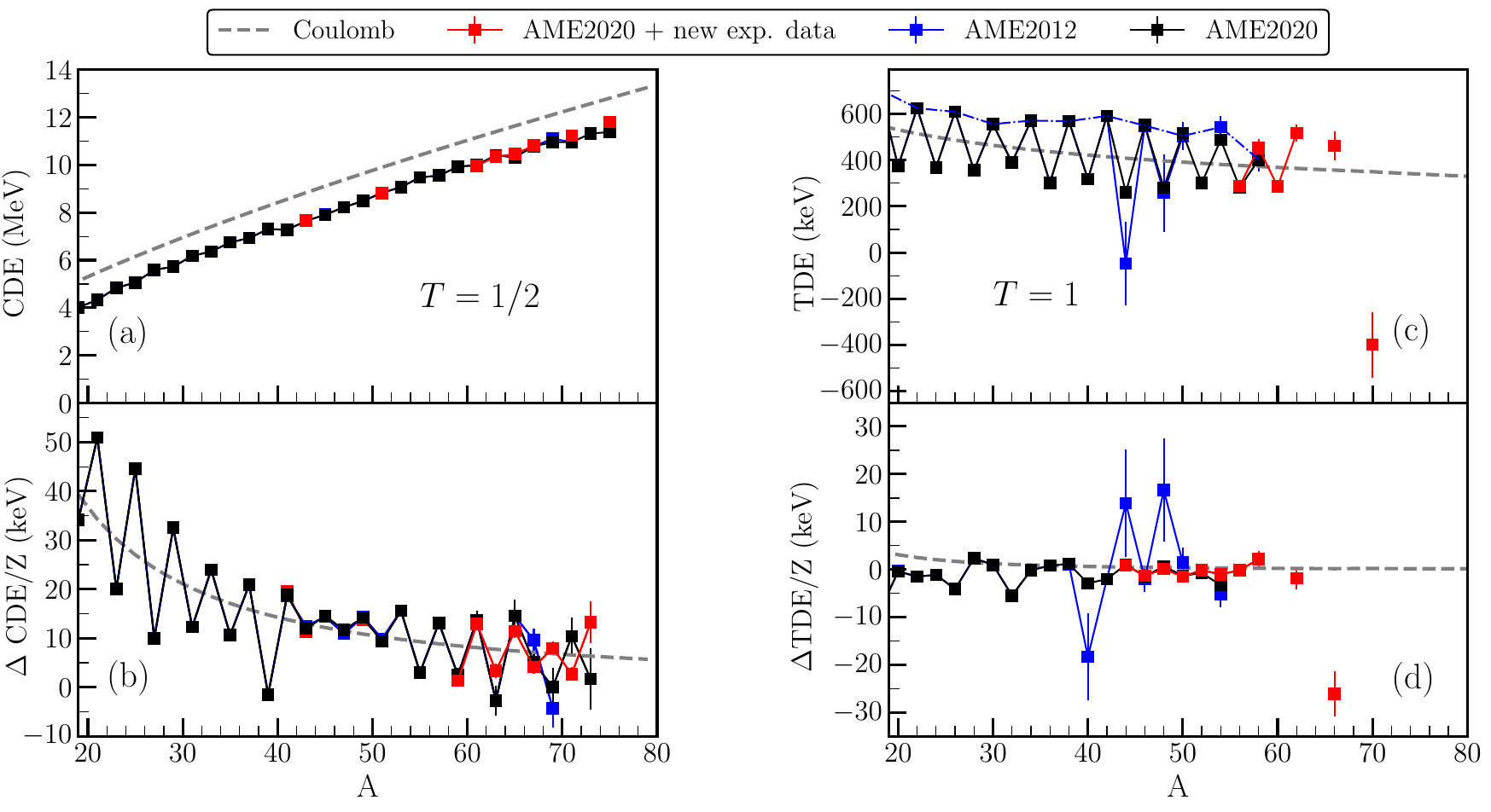}
\caption{\label{fig_cal_cde} Experimental (a) CDE, (b) CDE differences between $A$ and $A~+~2$ nuclei (shown as $\Delta$CDE/$Z$), (c) TDE, and (d) TDE differences between $A$ and $A + 4$ nuclei (shown as $\Delta$TDE/Z). For comparison, Coulomb energies calculated assuming nuclei as uniformly charge spheres\cite{bentley2007} are also shown.}
\vspace{-2pt}
\end{figure*}

In Fig.~\ref{fig:fig1}, CDEs and TDEs starting from $A = 20$ are illustrated together with a simplest estimate for the nuclear Coulomb energy by assuming nucleus as an uniformly charge sphere\,\cite{bentley2007}.
The corresponding values were  determined utilizing the masses from Atomic Mass Evaluation 2012 (AME2012)~\cite{ame2012}, and 2020 (AME2020)~\cite{ame2020}, as well as the latest experimental input. One observes in Fig.~\ref{fig:fig1}(a) that the calculated Coulomb energy generally overestimates  the  CDEs for $T = 1/2$. 
These large deviations from the data, by more than 1 MeV on average, were qualitatively explained  by the exchange effect due to the Pauli principle, which keeps the protons apart and thus weakens the Coulomb repulsion \cite{nolen1969}.
\minew{There have been many attempts to resolve this discrepancy known as the Nolen-Schiffer anomaly\cite{Shahnas1994,SUZUKI1992141,sagawa2024}.}

With the Coulomb energy as a mean value, an odd-even zigzag pattern of $\Delta$CDE/$Z$ is seen in Fig.~\ref{fig:fig1}(b). The pattern exhibits oscillations \minew{with} a common phase, but at $A = 69$, the mass data from both AME2012 and AME2020 suggest a phase reversion. This phase reversion was noticed in Ref.~\cite{kaneko2013} and discussed in detail in Ref.~\cite{tu2014}.
Notably, the shell model calculation in Ref.~\cite{kaneko2013}, no matter with or without the INC interaction in the $f_{7/2}$ shell, could not substantiate this perceived phase reversion. It was speculated~\cite{kaneko2013} that the ``anomaly" might be attributed to the cited mass value of $^{69}$Br in Ref. \cite{ame2012} being for an isomer rather than the ground state. In AME2020, with the proton separation energy determined from the $\beta$-delayed proton emission of $^{69}$Kr experiment~\cite{santoplb2014}, the mass excess of  $^{69}$Br was updated to be $-46260(40)$~keV, that is by 150~keV more bound than the one in AME2012. However, the phase revision at $A = 69$ still remains and the reversed phase persists for the heavier masses of $^{73}$Rb and $^{75}$Sr (see black squares in Fig.~\ref{fig:fig1}(b)). It is remarkable that by applying the new experimental mass values, the previously suggested phase revision at $A \geq 69$ disappears. The change is due to the new experimental mass excess of $^{71}$Kr of  $-46056(24)$~keV, which is by 270~keV less bound than the value in AME2020. Furthermore, the phase at $A = 73$ was rectified by the new mass of $^{75}$Sr, which is by 420~keV less bound than the AME2020 value derived from the $Q$-value of the $\beta$ decay of $^{75}$Sr \cite{huikari2003mirror}.

The TDEs for $T = 1$ starting from $A  = 20$ are shown in Fig.~\ref{fig:fig1}(c), and compared to the Coulomb energy estimate \cite{bentley2007}.
With the new mass data, the TDEs for $A = 4n + 2$ are conspicuously extended from $A = 58$ (see Ref.~\cite{kaneko2013}) to $A = 70$ and those for $A=4n$ are now presented up to $A = 60$. The estimated Coulomb energy serves as a smoothly changing average, around which TDEs oscillate, \yl{thereby dividing data into two branches either above or below it.}  Within error bars of the last few data points,  the TDEs of $A = 4n + 2$ ($A = 4n$) lie, on average, roughly by 150 keV above (below) the Coulomb energy prediction.

The blue dash-dotted-line in Fig.~\ref{fig:fig1}(c) connects the $A=4n+2$ TDEs determined by the AME2012 mass data and is shown to emphasize the sudden drop at $A$ = 58. 
This sudden drop seemed to suggest a termination of the odd-even oscillation in TDE \cite{kaneko2013}. 
Now with the updated mass data, the zigzag trend continues. This change is due to the updated masses of $^{54}$Ni and $^{58}$Zn. Compared to AME2012, the new mass of $^{54}$Ni is by 65~keV more bound and that of $^{58}$Zn is by 52~keV less bound, which result in a decrease in TDE at $A = 54$ and \yl{an} increase at $A = 58$, thus retaining the  smooth trend for overall TDEs of the $A = 4n + 2$ branch. 

A large, negative TDE at $A = 70$ is observed, which deviates strongly from the general trend in Fig.~\ref{fig:fig1}(c). This anomaly was discussed in detail by analyzing the systematics of $b$ and $c$ coefficients of the quadratic IMME~\cite{huangprc2023}, and was attributed to the presently adopted mass of $^{70}$Br in AME2020.

\begin{figure*}[t] \label{fig:fig2and4}
\includegraphics[width=1.7\columnwidth]{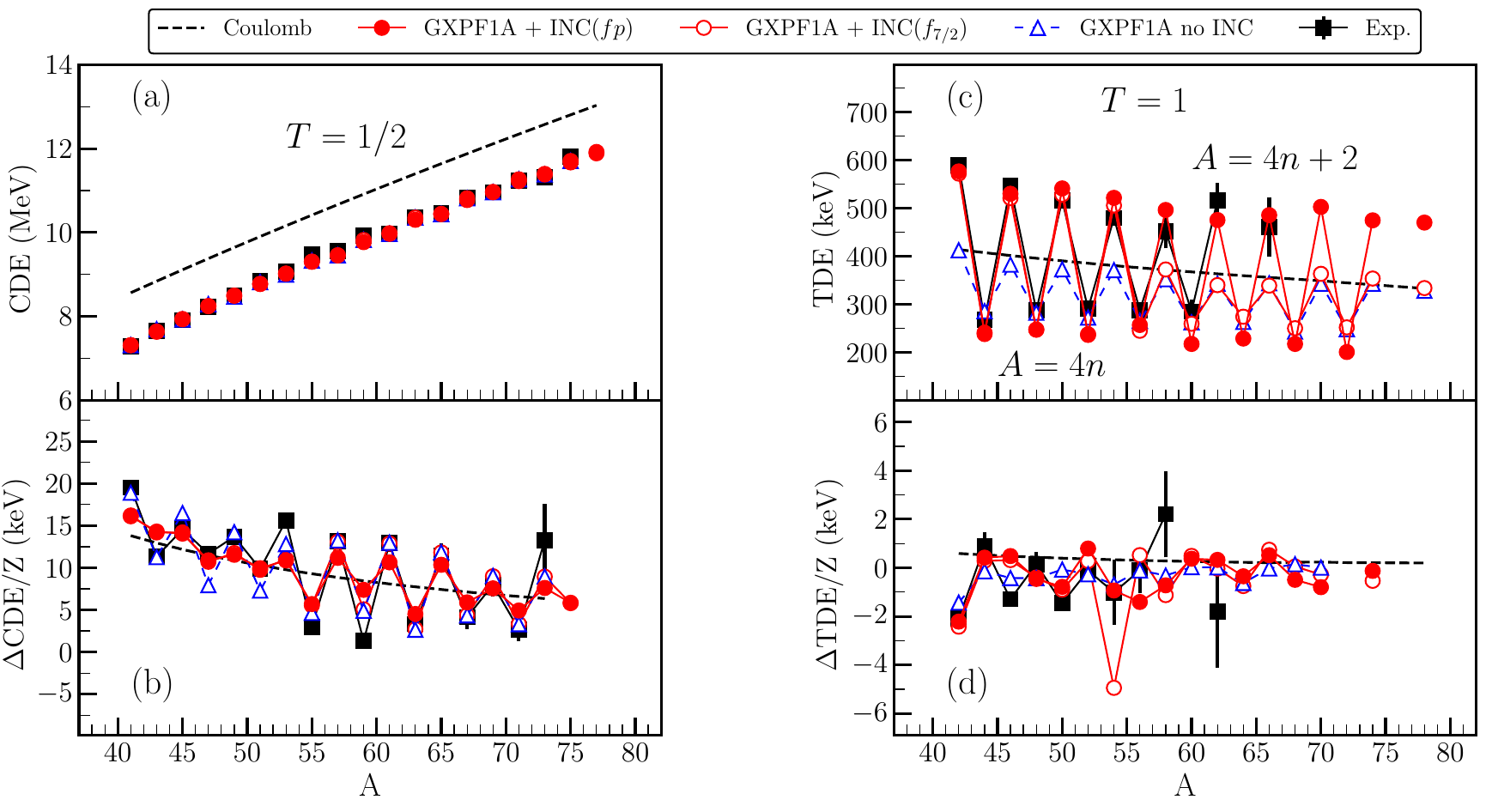}
\caption{ Experimental (a) CDE for $T=1/2$ and (b) CDE differences between $A$ and $A+2$ nuclei (shown as $\Delta$CDE/Z), (c) TDE for $T = 1$  and (d) TDE differences between $A$ and $A + 4$ nuclei  are compared with the shell-model calculation. For comparison, estimated Coulomb energies \cite{bentley2007} are also shown.}
\end{figure*}

The Coulomb estimate in Fig.~\ref{fig:fig1}(d) predicts very small  $\Delta$TDEs for lighter nuclei and the values approach zero with increasing $A$. Experimental data, with only small scatter in the light mass region, do not deviate much from this trend. Especially in the range of $A = 44-62$, the new $\Delta$TDE data follow precisely the Coulomb energy prediction. The previously-found irregular values at $A = 40, 44$ and 48, determined with the data from AME2012, now converged to the Coulomb estimate. Also the small drop at $A = 54$, due to the sudden drop in TDE at $A = 58$ when data of AME2012 were used, disappears.

The experimental CDE and TDE values determined with the new masses can be found in the supplementary material\,\cite{Supplement}. The data beyond $A = 58$, TDE for $T = 1$ and CDE for $T = 1/2$, permit us to extend the ISB study to the upper $fp$ shell.
We have performed shell-model calculations with the same nuclear INC interaction as in the $f_{7/2}$ orbit  \cite{kaneko2013}. Figures \ref{fig:fig2and4}(a) and (b) show the calculated CDEs for isospin $T=$ 1/2. The \yl{calculations reproduce} all experimental data reasonably well. The odd-even \yl{oscillation} in CDE between  $A$ and $A + 2$ \yl{nuclei}, described in $\Delta$CDE/$Z$, are reproduced correctly. The oscillations can be  understood by the difference between the proton- and neutron-pairing gaps due to the Coulomb force \cite{kaneko2013}. Nevertheless, except that the INC forces lead to a reduction in staggering amplitudes of CDEs for $A= 41 - 53$, no significant differences can be seen in the calculations with and without INC. Finally at $A = 73$, the present experimental value of  $\Delta$CDE/Z for  $T=1/2$ deviates from the calculation by approximately 6~keV (about 1.3 $\sigma$) which may suggest an  improvement of the mass of $^{75}$Rb.

In contrast,  significant effects due to the INC interaction are found in TDEs for $T=1$. As shown in Fig. \ref{fig:fig2and4}(c), for $A=$ 42 - 78, experimental TDE data indicate large oscillations between the $A = 4n + 2$ and $A = 4n$ \yl{data} branches. The calculated results (filled circles) with the INC interaction for the full $fp$ shell are in excellent agreement with the data. When the INC interaction is switched off, the TDEs for the $A=4n+2$ branch decrease drastically, while those for $A = 4n$ are  hardly affected. The distinct response to the INC force from the two branches can be interpreted as follows. As triplet nuclei in a TDE of $A = 4n+2$ branch can be understood as putting $pp$, $nn$, and $pn$ pairs on top of a common even-even $N = Z$ core, the $T = 1$ TDEs for $A = 4n + 2$ triplets are considered as good indicator of the isotensor force acting among them. As shown in Fig. \ref{fig:fig2and4}(c), the prediction of the isotensor force of 165 keV ($=\beta_{pp}+\beta_{nn}-2\beta_{pn}$, see Ref. \cite{Supplement}) agrees well with the  \yl{average} difference of $\sim$150 keV between the calculated TDEs with and without the INC interaction included. If the INC for the upper $fp$ shell is switched off, the TDEs drop suddenly at $A = 58$,  \yl{consistent} with the previous result in Ref.~\cite{kaneko2013}.

With the  calculated TDE value, 503~keV at $A =70$, together with the known masses of $^{70}$Se and $^{70}$Kr (see Ref. \cite{Supplement}), the mass excess of $^{70}$Br can be deduced to be $-51877(70)$~keV.
This value agrees with the prediction \yl{from the systematic study\,\cite{huangprc2023}} of $-51934(16)$~keV. 

We can thus conclude that the INC interaction is significant for $A = 4n+  2$ TDEs, and the inclusion of INC in the upper $fp$ shell in shell-model calculations turns out to be as important as in the $f_{7/2}$ orbit. The TDE for $T=$ 1 is a good indicator of the isotensor INC force, which is estimated to be 150 keV from the present study in the $A =$ 42 -78 region.

Figure \ref{fig:fig2and4}(d) shows the calculated $\Delta{\rm TDE}/Z$ for both branches of data points. In consistence with the data, they are close to zero and vary only slightly in the entire mass region.
The $\Delta$TDE/$Z$ for $A = 4n + 2$ may be regarded as another quantity to underline the (nearly) $A$-independent upward shift of TDEs due to the INC effect.

Little is currently known about the INC effects in the  heavier mass regions toward $^{100}$Sn. An initial investigation of CDEs and TDEs was carried out   with the JUN45 interaction in the $fpg$ model space including the $g_{9/2}$ orbit \cite{kaneko2013}. The study was incomplete because no INC forces were included. 
Based on the present analysis, we may expect that the INC effects would still be important and the discussed TDE oscillations in this work would continue  \yl{to} nuclei in the  $A= 80-100$ region.

In summary, the new mass data allow us, for the first time, to make a significant step toward  understanding the isospin symmetry breaking in the upper $fp$ shell. For CDEs, the regular odd-even staggering phase of the CDE differences  persists up to $A = 73$, rectifying the previously-found phase inversion at $A = 69$. For TDEs, a general odd-even staggering pattern up to $A = 62$ was observed. The new data refuted several ``anomalies" in TDEs suggested before.  Detailed shell-model calculations, employing one uniform nuclear INC interaction for the entire $fp$ shell, provided an excellent description  \yl{of} the experimental CDE and TDE values. 
The largest change in TDE due to the INC effect for  $A = 4n + 2$ triplet nuclei, having $pp$, $nn$, and $pn$ pairs on top of a common even-even $N = Z$ core, is estimated to be 150 keV for the isotensor component of  isospin violating interactions. Finally, the astonishingly-regular behavior of the charge symmetry and charge independence of the
attractive $NN$ interactions discussed here may justify an extension of the INC concept up to the $^{100}$Sn region.

\begin{acknowledgments}
This work is supported in part by the National Key R\&D Program of China (Grant No. 2021YFA1601500), the Strategic Priority Research
Program of Chinese Academy of Sciences (Grant No.
XDB34000000), the Youth Innovation Promotion Association
of the Chinese Academy of Sciences (Grant No. 2022423), and the NSFC
(Grant Nos. 12135017,  12121005,  11975280,  12322507, 12305151, and 12235003).
H. F. L. acknowledges the support from the Special Research Assistant Project of the Chinese Academy of Sciences.
\end{acknowledgments}



\bibliography{cde_tde_paper}

\end{document}